\documentclass[twocolumn]{aa}
\usepackage{amssymb}

\def\al{&\!\!\!\!}

\def\f{\frac}
\begin{document}

\title{(Erratum) An $f(R)$ gravitation for galactic environments}

\author{R. Saffari and Y. Sobouti}
\institute{Institute for Advanced Studies in Basic Sciences, P. O.
Box 45195-1159, Zanjan,
      Iran\\   }

\date{ }
\abstract {Without abstract \keywords{ cosmology: theory --
cosmology: dark matter -- gravitation -- galaxies: photometry  } }

\authorrunning{R. Saffari}
\titlerunning{An $f(R)$ gravitation for galactic environments}

\maketitle


We regret that an initially typographical error in the geodesic
equations, Eq. (15) of \cite{Sob07}, has propagated through the
paper. Corrections are as follows:
\begin{eqnarray}
\al\al\frac{d^2r}{d\tau^2}+\frac{1}{2}\frac{A'}{A}\left(\frac{dr}{d\tau}\right)^2-
\frac{r}{A}\left(\frac{d
\varphi}{d\tau}\right)^2+\frac{1}{2}\frac{B'}{A}\left(\frac{dt}{d\tau}\right)^2=0,
\hspace{0.13cm}~(15)\nonumber
\end{eqnarray}
\begin{eqnarray}
\al\al \f{J^2}{r^3}=\f{1}{2}\f{B'}{B^2},
\hspace{6.24cm}~~(20)\nonumber
\end{eqnarray}
\begin{eqnarray}
\al\al v^2=\f{1}{2}\f{r
B'}{B}=\f{1}{2}(\alpha-\f{rA'}{A}),\hspace{3.987cm}~~(22)\nonumber
\end{eqnarray}
\begin{eqnarray}
\al\al v^2=\frac12\alpha+\frac12(1-\frac12\alpha)
\frac{{(\frac{s}{r})}^{1-\alpha/2}+2\lambda{(\frac{r}{s})}^{2(1-\alpha/2)}}
{1-{(\frac{s}{r})}^{1-\alpha/2}+\lambda{(\frac{r}{s})}^{2(1-\alpha/2)}},\hspace{0.25cm}~~(23)\nonumber
\end{eqnarray}
\newpage
\begin{eqnarray}
\al\al v^2=\frac{1}{2}\alpha c^2
+\frac{GM}{r}\left[1-\frac{1}{2}\alpha\left\{1+\ln\left(\frac{2GM
}{c^2r}\right)\right\}\right].\hspace{0.695cm}~~(24)\nonumber
\end{eqnarray}
Numbers following each equation are the equation numbers of
\cite{Sob07}.

In table 1, column 5, $2(v_{\infty}/c)^2\times10^{-7}$ and in
column 6, $\alpha_0\times10^{-7}$ are the correct headings. For
\textit{NGC~801} the asymptotic velocity, $v_{\infty}$ is
$208~km/s$, corresponding to
$\alpha_0=2.71\times10^{-7}$. The changes are minute and in no way
alter the conclusions of the paper.

\end{document}